# A Study on the Factors That Influence the Consumers' Trust on E-commerce Adoption

Yi Yi Thaw
Department of Computer and Information Sciences
Universiti Teknologi PETRONAS,
Tronoh, Perak, Malaysia

Ahmad Kamil Mahmood
Department of Computer and Information Sciences
Universiti Teknologi PETRONAS,
Tronoh, Perak, Malaysia

P.Dhanapal Durai Dominic
Department of Computer and Information Sciences
Universiti Teknologi PETRONAS,
Tronoh, Perak, Malaysia

*Abstract*—**The development of electronic commerce is characterized with anonymity, uncertainty, lack of control and potential opportunism. Therefore, the success of electronic commerce significantly depends on providing security and privacy for its consumers' sensitive personal data. Consumers' lack of acceptance in electronic commerce adoption today is not merely due to the concern on security and privacy of their personal data, but also lack of trust and reliability of Web vendors. Consumers' trust in online transactions is crucial for the continuous growth and development of electronic commerce. Since Business to Consumer (B2C) e-commerce requires the consumers to engage the technologies, the consumers face a variety of security risks. This study addressed the role of security, privacy and risk perceptions of consumers to shop online in order to establish a consensus among them. The analyses provided descriptive frequencies for the research variables and for each of the study's research constructs. In addition, the analyses were completed with factor analysis and Pearson correlation coefficients. The findings suggested that perceived privacy of online transaction on trust is mediated by perceived security, and consumers' trust in online transaction is significantly related with the trustworthiness of Web vendors. Also, consumers' trust is negatively associated with perceived risks in online transactions. However, there is no significant impact from perceived security and perceived privacy to trust in online transactions.**

***Keywords-perceived security and perceived privacy; perceived risk; trust; Web vendors; consumer behavior.***

## I. INTRODUCTION

This study focuses on the aspect of e-commerce that utilizes the Internet and World Wide Web (WWW) as the technological infrastructure to communicate, distribute and conduct information exchange that would consequently lead to the commercial transactions between Web vendors and consumers. In addition, this study would likely to identify the main security and privacy issues concerns and the trustworthiness of the Web vendors to engage in e-commerce transaction and the effectiveness of security methods and applications in ensuring the confidentiality, integrity and privacy of e-commerce transactions. The present research intends to identify the factors which are directly related to consumers' trust to adopt e-commerce in Malaysia. Therefore, this study is undertaken to answer the following research questions: Do consumers' security and privacy concerns of online transaction significantly relate to their trust in e-commerce adoption? How do the trustworthiness and reliability of the Web vendors relate to the consumers' adoption of e-commerce? What are the inter-relationships of security and privacy concerns, trust beliefs and risk perception, and how do these factors affect consumers' behavior intention to adopt e-commerce?

## II. LITERATURE REVIEW

E-commerce has gained considerable attention in the past few years, giving rise to several interesting studies and industrial application, due to the Internet has created enormous change in the business environment. The Malaysian Government has made a massive move by launching the Multimedia Super Corridor (MSC) whereby one of its seven flagship applications includes the active promotion of the electronic business activities in the country. However, the acceptance level of the electronic commerce by the Malaysian consumers is still regarded very low compared to the other parts of the world especially the developed countries like the United States and the European Union. For example, the Small- and Medium-Sized Industries Association of Malaysia said in late 2005 that less than 5% of its members were involved in B2C business. According to Krishnan [1], the majority of Malaysians interested in e-commerce are males (66%) and males below 30 years (42%) is the largest individual group of Malaysians interested in e-commerce.

Considerable numbers of research findings [2], [3] and [4] have indicated that although e-commerce is spreading worldwide, customers are still reluctant to deal with it because of the security and privacy issues. A study of consumer-perceived risk in e-commerce transactions by Salam *et al.* [5] indicated that consumers simply do not trust online vendors to

Universiti Teknologi PETRONAS, MALAYSIA

engage in transactions involving money and personal information. According to the authors, consumer-perceived risk is reduced with the increase in institutional trust and economic incentive.

Ahmed *et al.* [6] surveyed that the major concerns on e-commerce adoption in Malaysia are: security and privacy over online transaction process and trust and reliability of online vendors. They suggested that in order to be successful in electronic marketplace, the organizations are expected to expend their resources and exert efforts to ensure that consumers' concerns are adequately addressed. Dauda *et al.* [7] studied the perceived e-commerce security influence on adoption of Internet banking, and the role of national environmental factors such as attitude, subjective norms, and perceived behavioral control factors towards adoption, and compares these factors with Singapore Internet banking adoption. They found that consumer perceived non-repudiation, trust relative advantage Internet experience and banking needs are the most important factors that affect adoption in Malaysia. Organizations were reluctant to use e-commerce as they felt that the transactions conducted electronically were open to hackers and viruses, which were beyond their control. Khatibi and Scetharaman [8] mentioned that Malaysian e-commerce industry has not taken off as expected. By means of a survey of 222 Malaysian manufacturers, traders and service providers, the authors concluded that from the company's point of view, the main barriers of e-commerce adoption are: concern on security and privacy followed by the hustle of keeping up with the technology, uncertainties regarding rules and regulations, high set up cost of Ecommerce, lack of skilled workers and so on. The authors suggest that any policy that aims at promoting e-commerce should take these factors into consideration.

According to mid-2005 survey conducted by the Malaysian Communications Multimedia Commission (MCMC), only 9.3% of internet users had purchased products or services through the internet during the preceding three months [9]. The primary reasons cited for this are: lack of security and privacy of consumers' personal data including credit card number, identity theft, virus, break-in attacks, denial-of-service, and so on. Lincoln Lee [10], Senior Analyst, Telecommunication Research, IDC Malaysia, mentioned that "the Malaysia ecommerce market has exhibited a healthy growth rate of 70% in 2006 in comparison with that in 2005. However, in order to ensure sustainable growth, there is still plenty of work to be done to develop this industry into a mature market". Jawahitha [11] raised serious concern on the protection of Malaysian consumers dealing with e-commerce transactions. According to her, the existing laws pertaining to conventional businesses are not sufficient to address the issues in e-commerce. Malaysian government has already taken steps to pass new laws and to amend some of the existing laws and until this effort is materialized, the Malaysian electronic consumers would not be having adequate protection. To protect e-commerce consumers' privacy, Malaysian legislators have devised a personal data protection bill [12]. The author examined the nature, manner and scope of personal data protection under this Bill. She suggests that instead of being concerned with the full range of privacy and surveillance issues, the Bill deals only with the way personal data is collected, stored, used and accessed.

In essence, numerous research papers have been published during the last few years on various issues pertaining to e-commerce. Since this paper deals with building consumers' trust in e-commerce transaction, it only cites literature relevant to the issue. The present research is intended to fill-up the gap on Malaysian consumers regarding identification of factors that help build their trust in greater e-commerce participation.

## III. RESEARCH DESIGN AND METHOD

The main objective of this study is to identify the factors that contribute to the consumers' willingness to engage in e-commerce transactions, and further study the relationship between those factors. Therefore, this study will focus on the following sub-objectives:

- To study whether or not consumers' perceived security and privacy of online transaction significantly affect their confidence to adopt e-commerce.
- To identify the factors of trust with web vendors to engage in transactions involving money and personal data.
- To study the role of institutional trust and economic incentive in consumers' perceived risk in the context of e-commerce.

The factors considered to be influencing consumers' confidence to adopt e-commerce are grouped into four main categories: consumers' attitudes towards secure online transaction processing systems, privacy of consumers' personal data, trust and reliability of online vendors, and consumers' perceived risk in e-commerce transactions. The model to be tested is shown in Figure 1.

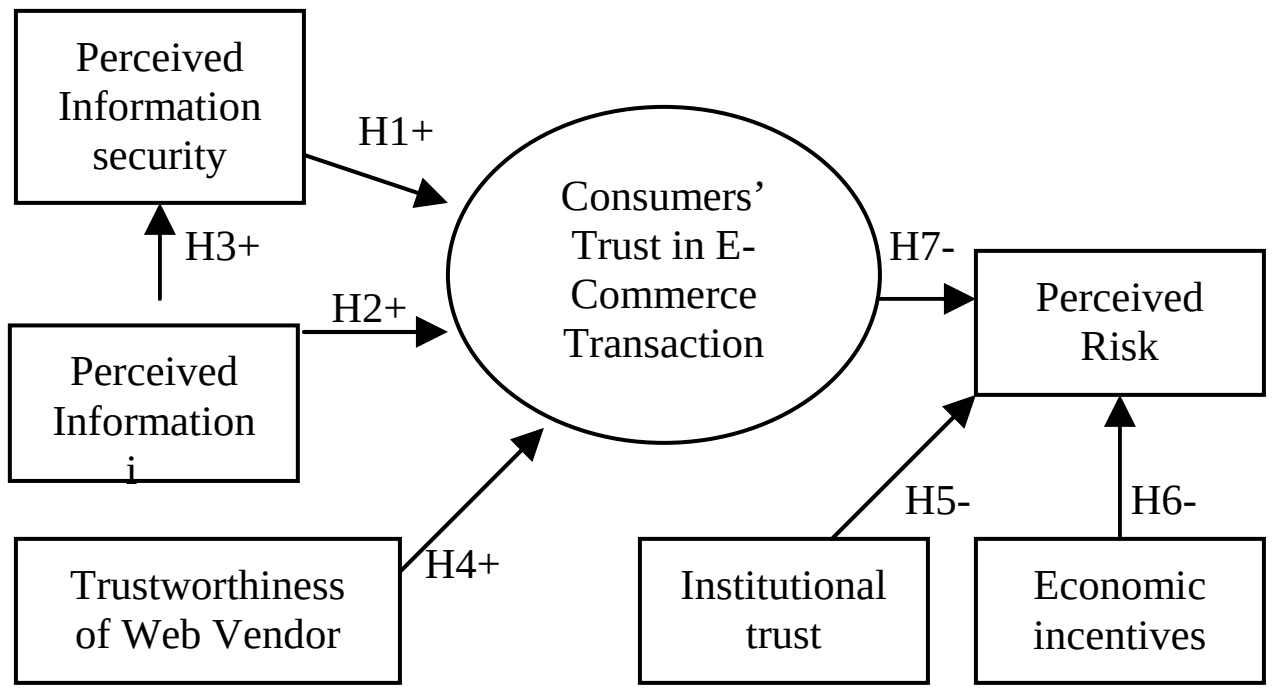

**Figure 1. Research Model.**

Specifically, the following hypotheses are to be tested:

**H1:** A consumer's perceived security of online transaction positively contributes to his/her trust in online transaction.

**H2:** A consumer's perceived privacy of online transaction positively contributes to his/her trust in online transaction.

**H3:** The influence of a consumer's perceived privacy of online transaction on trust is mediated by perceived security.

**H4:** A consumer's trust in online transaction is positively related with the trustworthiness of Web vendor.

**H5:** The increase in institutional trust reduces consumers' perceived risk in online transaction.

**H6:** The increase in economic incentives reduces consumers' perceived risk in online transaction.

**H7:** A consumer's trust in online transaction is negatively associated with perceived risk in online transaction.

A survey instrument in the form of questionnaire was developed that consisted of three sections. Section 1 consisted of questions to collect respondents' personal information (i.e., gender, age, race, etc). Section 2 consisted of questions covering some of the variables related to online purchase and adoption of electronic commerce. Specifically, the questions were designed to collect information on frequency of internet use, frequency of online purchases, intention to continue online purchasing, etc. Section 3 consisted of questions covering some of the variables related to factors affecting e-commerce security, privacy, and trust as well as risk perceptions. Questions in this section collected information related to attitudes towards secure online transaction processing system, privacy of personal data, trustworthiness of Web vendors, and consumers' perceived risk. All the variables in this section employed Likert scale with endpoints ranging from 1 (strongly disagree) to 5 (strongly agree).

Before sending the questionnaires to the mass, it was pre and pilot tested through a series of informal interviews with faculty and doctoral students to ensure that they were properly operationalized. The items measures were suitably modified or adapted from extant literature. Based on pilot study with 25 master and doctoral students for comprehensiveness, clarity and appropriateness, 5 items for perceived security, 6 items for perceived privacy, 5 items for trustworthiness of Web vendors, 3 items for consumers' perceived risk, 2 items for economic incentive, 2 items for institutional trust and 2 items for consumers' trust were incorporated into the study instrument. In this survey, the target group of respondents were the internet savvy students. 85 full-time final year undergraduate students (50.6% males and 49.4% females) from two local universities are participated in this study. The majority of the respondents (about 98.8%) are age between 20 to 30 while remaining about 1.2% is age between 31 to 40. In term of races, about 57.6% are Malay while about 18.8% are Chinese and about 15.3% are Indian.

## IV. Data Analysis

Out of the 85 respondents, almost all the respondents (about 96.5%) report that they frequently use the internet while the remaining 3.5% seldom use the internet. The respondents did not have experience in online purchases and they were asked about the possibility of their willingness to make online purchases in the near future. About 49.4% are not willing to purchase in the near future and about 8.3% are willing to make online purchases in future. Furthermore, the respondents who are not willing to purchase online in the near future were asked about the reason(s) for that. The major reason (35.5%) was cited to be the concern on security and privacy of their personal data, followed by lack of interaction (about 27.1%) and cannot feel product (about 22.4%). All the respondents were also asked about their opinion on credit card security for online purchases. The majority of the respondents (about 54.1%) believe that the use of credit card for online purchases is not safe, while about 11.8% believe somewhat safe. About 8.2% of the respondents are indifferent on online credit card security and the remaining (about 24.7%) respondents are not sure about this.

### A. *Descriptive Analysis*

*1) Information security concerns:* Regarding online information security concerns, only 10.6% of the respondents agree that they would feel totally safe providing sensitive information about themselves over the Web while majority (about 57.7%) of the respondents do not believe this, and about 31.8% of the respondents remained neutral on this question. On the online payment, about 22.4% of the respondents agree that the payment information they enter online is safe and accessible only by the intended persons while majority (about 41.1%) of the respondents do not believe this. The remaining 36.5% of the respondents remained indifferent to the question. On the integrity of the online transactions, only 11.8% of the respondents believe that the information they enter online is not altered in transit while 33.0% of the respondents do not believe this. The remaining majority (about 55.3%) of the respondents remained neutral on this question. About 17.6% of the respondents agree that they would not hesitate to make purchase from the Web because of security issues of sensitive information and about 40.0% of the respondents do not agree this. The remaining 42.4% of the respondents remained indifferent to the question. Overall, about 31.8% of the respondents believe that there is an adequate control in place to ensure security of personal data transmitted during online transaction processing while about 30.6% of the respondents do not believe this, and about 37.6% of the respondents remained neutral on this question.

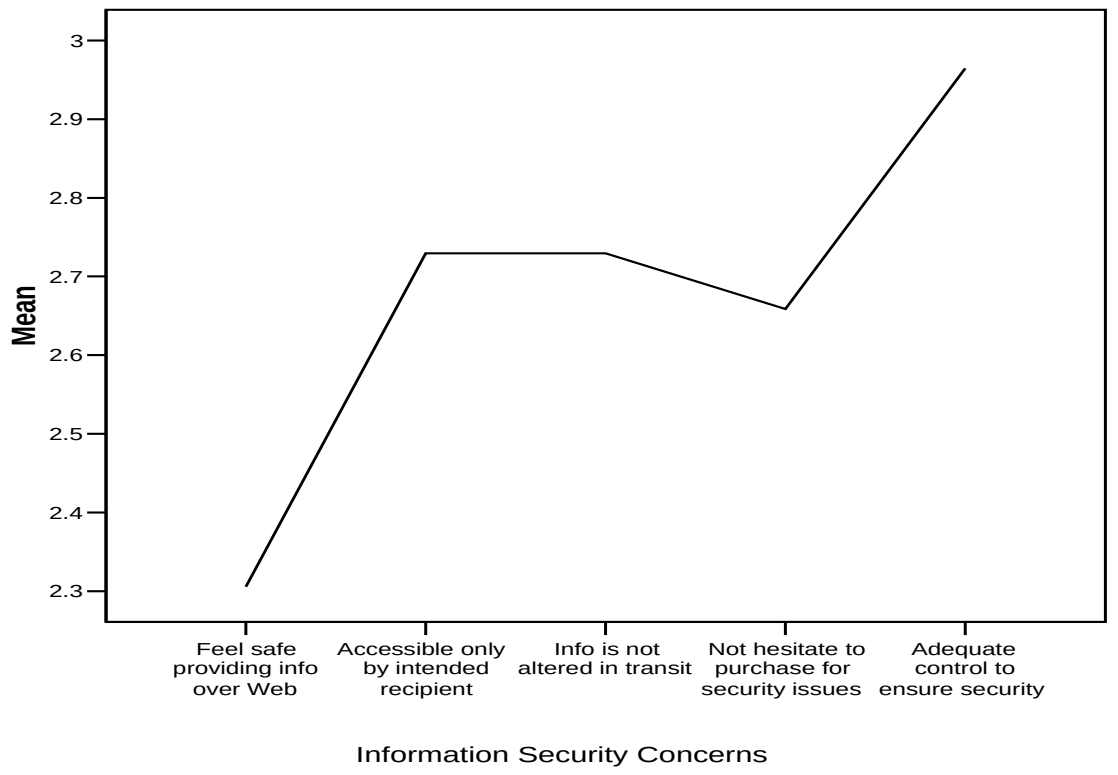

**Figure 2. Mean of Information Security Concerns.**

*2) Information privacy concerns:* Regarding the information misused, about 36.5% of the respondents believe that their personal information would not be misused when transacting with online companies and about 23.5% of the respondents do not believe this. The remaining 40.0% of the respondents remained neutral on the question. Regarding the control over information, about 42.3% of the respondents believe that they have control over how the information they provide will be used by online companies while about 24.7% of the respondents do not believe this. The remaining 32.9% of the respondents remained indifferent to the question. Moreover, about 31.8% of the respondents believe that they can later verify the information they provide during a transaction with online companies while about 24.7% of the respondents do not believe this. The remaining 43.5% of the respondents remained neutral on the question. In addition, only 25.9% of the respondents believe that online companies will not reveal their sensitive information without their consent while about 30.6% of the respondents do not believe this, and majority (about 43.5%) of the respondents remained neutral on this question. Regarding the effective mechanism, about 35.3% of the respondents believe that there is an effective mechanism to address any violation of the sensitive information they provide to online companies while about 20.0% of the respondents do not believe this. The remaining majority (about 44.7%) of the respondents remained indifferent to the question. Overall, about 35.3% of the respondents believe that there is an adequate control in place to protect the privacy of personal information within online companies while about 18.8% of the respondents do not believe this, and majority (about 45.9%) of the respondents remained indifferent to this question.

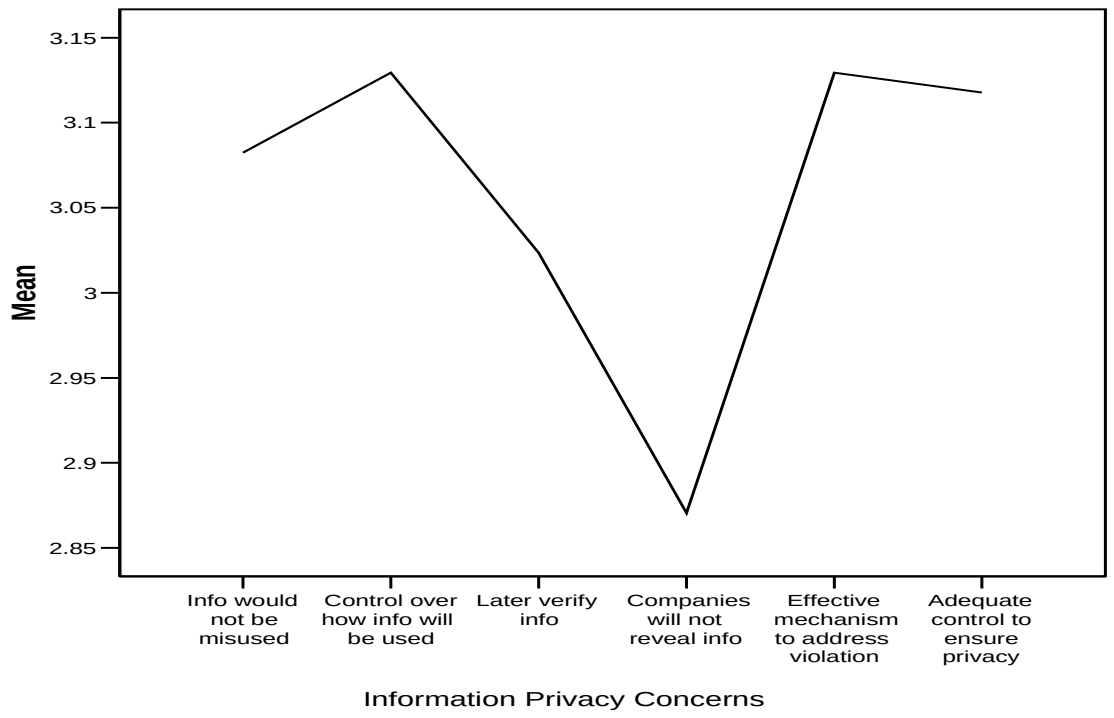

**Figure 3. Mean of Information Privacy Concerns.**

*3) Trustworthiness of Web Vendors:* Regarding trust beliefs of Web vendors, about 36.5% of the respondents believe that online companies will act with high business standards while about 24.7% of the respondents do not believe this. The remaining 38.8% of the respondents remained indifferent to the question. On the skills and expertise, majority (about 48.2%) of the respondents believe that online companies have the skills and expertise to perform transactions in an expected manner and about 22.3% of the respondents do not believe this. The remaining 29.4% of the respondents remained neutral on the question. Regarding whether online companies are dependable, about 30.6% of the respondents believe that online companies are dependable while about 24.7% of the respondents do not believe this. The remaining 44.7% of the respondents remained indifferent to the question. Moreover, about 29.4% of the respondents believe that online companies do not have ill intensions about any of their consumers while about 31.7% of the respondents do not believe this. The remaining 38.8% of the respondents remained indifferent to the question. Overall, only 22.4% of the respondents believe that online companies are trustworthy while about 25.9% of the respondents do not believe this, and majority (about 51.8%) of the respondents remained neutral on this question.

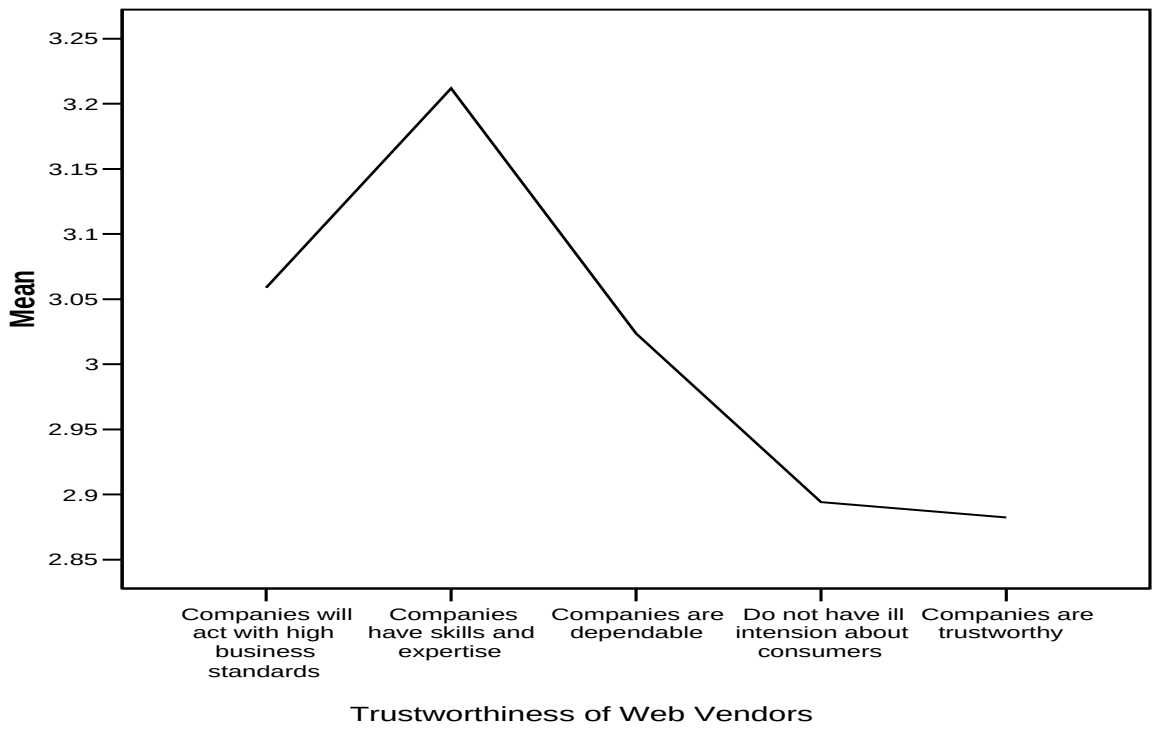

**Figure 4. Mean of Trustworthiness of Web Vendors.**

*4) Risk perception:* Regarding risk perception, majority (about 48.3%) of the respondents believe that providing credit card information over the Web is unsafe while only 18.8% of the respondents do not believe this. The remaining 32.9% of the respondents remained indifferent to the question. In addition, majority (about 54.1%) of the respondents believe that it would be risky to give personal information to online companies while about 17.7% of the respondents do not believe this. The remaining 28.2% of the respondents remained indifferent to the question. Furthermore, majority (about 51.7%) of the respondents agree that there would be too much uncertainty associated with providing personal information to online companies and about 18.8% of the respondents do not agree on this. The remaining 29.4% of the respondents remained neutral on this question.

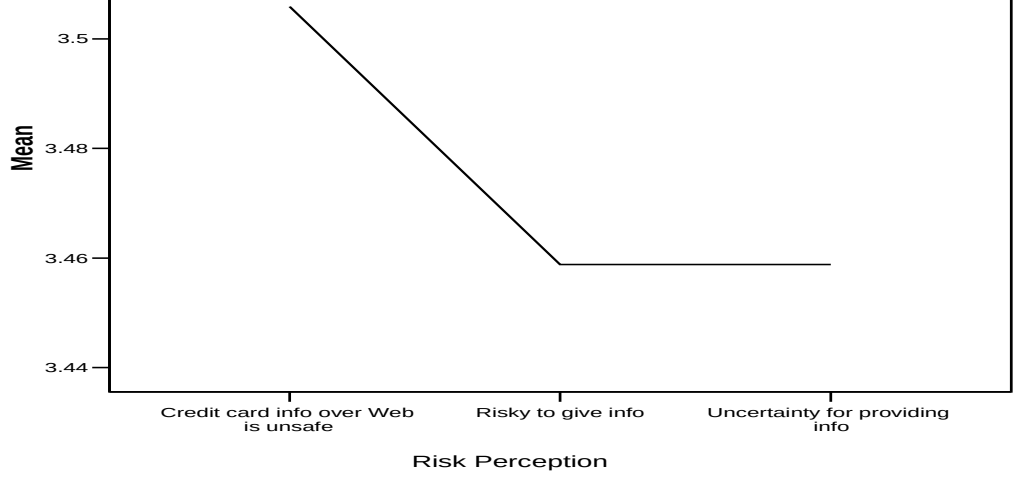

**Figure 5. Mean of Risk Perception.**

*5) Institutional trust:* Regarding institutional trust, about 55.3% of the respondents would trust to open financial account with a Bank, while about 17.6% do not agree on this and the remaining (about 27.1%) of respondents are not sure about this. Moreover, about 40.0% of the respondents would trust to open financial account with a major credit card company and only 8.2% of the respondents do not agree on this. The remaining majority (about 51.8%) of the respondents remained neutral on this question.

*6) Economic incentive:* Regarding economic incentive, about 24.7% of the respondents agree that providing credit card information over the Web would not matter much if the prices were considerably lower while about 38.8% of the respondents do not believe this. The remaining 36.5% of the respondents remained indifferent to the question. Furthermore, about 28.3% of the respondents agree that providing credit card information over the Web would not matter much if the products/services were of a higher quality and about 29.4% of the respondents do not agree on this. The remaining majority (about 42.4%) of the respondents remained neutral on this question.

*7) Consumers' trust:* On the use of more complex and advanced method, the majority (about 64.7%) of the respondents agree that their confidence to purchase online will be increased when more complex and advanced method is used to address concerns on their security and privacy while only 9.5% do not agree on this and 25.9% of the respondents remained neutral on this question. On providing all necessary guarantees to increase consumers' confidence to adopt e-commerce, the majority (about 58.8%) of the respondents agree that their confidence to adopt e-commerce will increase when online vendors provide all necessary guarantees to ensure security and privacy of their personal information, while only 8.2% do not agree on this and the remaining (about 32.9%) of respondents are not sure about this.

## *B. Reliability Analysis*

Reliability analysis was performed to assess the reliability of the scale used to measure the variables of interest. Reliability assessment of the entire scale was first computed followed by the assessments of individual items supposed to measure the research constructs. The overall reliability assessment of the entire scale was observed to be good with a Cronbach's alpha of 0.820**.** A second test was conducted to assess the degree to which the items on the scale measure the hypothesized research constructs. A total of twenty five items measuring six constructs and one dependent variable (five items measuring Information Security Concerns, six items measuring Information Privacy Concerns, five items measuring Trust Beliefs of Web Vendors, three items measuring Risk Perception, two items measuring Economic Incentive, two items measuring Institutional Trust, and two items measuring Consumers' Trust in e-commerce transactions) were assessed for reliability (See Table 1).

TABLE I. RELIABILITY ANALYSIS RESULTS

| **Items in the scale** | **Means and SD** | **Alpha** |
|---|---|---|
| Information Security Concerns | 13.39 (3.03) | 0.725 |
| Information Privacy Concerns | 18.35 (3.12) | 0.636 |
| Trust Beliefs of Web Vendors | 15.07 (2.91) | 0.660 |
| Risk Perception | 10.42 (2.20) | 0.642 |
| Economic Incentive | 6.05 (1.37) | 0.660 |
| Institutional Trust | 6.99 (1.63) | 0.779 |
| Consumers' Trust | 7.61 (1.66) | 0.707 |

## *C. Factor Analysis*

Factor analysis (factor extraction as well as varimax factor rotation) was conducted to identify the underlying factors affecting consumers' trust in e-commerce transactions. Considering all the 25 items on security and privacy of consumer's personal information, trust and reliability of web vendors, consumer's perceived risk, economics incentive, and institutional trust were analyzed using principal component analysis. "Total Variance Explained" shows the extent to which total variance of the observed variables is explained by each of the principal components. Initial factor extraction revealed seven components with an absolute magnitude of eigenvalue greater than 1.0. All the seven principal components together accounted for 63.596% of the total variance in the original 25 items. The results of factor analysis are shown in Table 2.

TABLE II. RESULTS OF FACTOR EXTRACTION AND FACTOR LOADING

| **Items** | **F1** | **F2** | **F3** | **F4** | **F5** | **F6** | **F7** |
|---|---|---|---|---|---|---|---|
| SE | 0.749 | | | | | | |
| EM | 0.745 | | | | | | |
| LVI | 0.694 | | | | | | |
| BS | 0.536 | | | | | | |
| ES | 0.538 | | | | | | |
| NHP | | 0.735 | | | | | |
| INA | | 0.687 | | | | | |
| NII | | 0.660 | | | | | |
| IR | | 0.534 | | | | | |
| DP | | | 0.669 | | | | |
| NRI | | | 0.629 | | | | |
| EP | | | 0.569 | | | | |
| COI | | | 0.527 | | | | |
| RGI | | | | -0.780 | | | |
| CCU | | | | -0.652 | | | |
| UC | | | | -0.619 | | | |
| TC | | | | | 0.861 | | |
| TB | | | | | 0.792 | | |
| NG | | | | | | 0.690 | |
| AM | | | | | | 0.548 | |
| HQ | | | | | | | 0.873 |
| LP | | | | | | | 0.514 |

Skills and expertise (SE) Effective mechanism (EM)
Later verify Info (LVI) Business standards (BS)
Ensure security (ES) Not hesitate to purchase (NHP)
Info not altered (INA) No ill intension (NII)
Intended recipient (IR) Dependable (DP)
Not reveal Info (NRI) Ensure Privacy (EP)
Control over Info (COI) Risky to give Info (RGI)
Credit card unsafe (CCU) Uncertainty (UC)
Trust with company (TC) Trust with bank (TB)
Necessary guaranties (NG) Advance Method (AM)
Higher quality (HQ) Low price (LP)

Most items loaded onto the extracted factors except from the some items that were conceptualized to measure the information security concerns, information privacy concerns and trust beliefs of web vendors. Item on adequate control to ensure security fairly loaded onto the factor of trust beliefs of web vendors, while the item on companies do not have ill intention about consumers slightly loaded onto the information security concerns factor. However, items on later verify info and effective mechanism to address violation of the information privacy concerns factor fairly loaded onto factor one (trust beliefs of web vendors). Also item on companies are dependable of the trust beliefs of web vendors factor loaded onto factor three. The tree items, namely, feel safe providing information over Web, information would not be misused, and companies are trustworthy had factor loading lower than 0.50.

### D. Hypothesis Testing

Pearson correlation coefficients were computed in order to test the relationships between each factor and consumers' trust in e-commerce transactions.

**H1:** *A consumer's perceived security of online transaction positively contributes to his/her trust in online transaction.*

The correlation coefficient between consumers' attitude towards secured online transaction and their confidence to adopt e-commerce was found to be with p = 0.545. Therefore, the research hypothesis is not accepted.

**H2:** *A consumer's perceived privacy of online transaction positively contributes to his/her trust in online transaction.*

The results of the study show that perceived privacy negatively affects the consumer's confidence to adopt e-commerce. The relationship is observed to be r=0.002 with p = 0.986. Therefore, we reject the research hypothesis.

**H3:** *The influence of a consumer's perceived privacy of online transaction on trust is mediated by perceived security.*

The results of the study show that consumer's perceived privacy of online transaction on trust is mediated by perceived security (r = 0.424). The relationship is observed to be statistically significant with significance level less than 0.01 (p = 0.000). Therefore, we accept the research hypothesis.

**H4:** *A consumer's trust in online transaction is positively related with the trustworthiness of Web vendor.*

The correlation coefficient between the trustworthiness of Web vendor and consumers' confidence to adopt e-commerce was found to be 0.218 with p = 0.045. Therefore, the research hypothesis is accepted.

**H5:** *The increase in institutional trust reduces consumers' perceived risk in online transaction.*

We found that the increase in institutional trust does not reduce a consumers' perceived risk in online transaction. The relationship is observed to be r = 0.148 with p = 0.176. Therefore, we reject the research hypothesis.

**H6:** *The increase in economic incentives reduces consumers' perceived risk in online transaction.*

We also found that the increase in economic incentives does not reduce a consumers' perceived risk in online transaction. The relationship is observed to be with p = 0.484. Therefore, we reject the research hypothesis.

**H7:** *A consumer's trust in online transaction is negatively associated with perceived risk in online transaction.*

The results of the study show that a consumer's trust in online transaction is negatively associated with perceived risk in online transaction. (r = 0.388). The relationship is observed to be statistically significant with significance level less than 0.01 (p = 0.000). Therefore, we accept the research hypothesis.

## V. Managerial Implications

The present study confirms that while consumers' perceived security directly acts upon trust in electronic commerce transactions, consumers' perceived privacy's effect on trust is mediated by perceived security. Those organizations that are involved in e-commerce as well as will be involved in e-commerce are expected that to act with high business standards and to have the skills and expertise to perform transactions in an expected manner. In addition, organizations should implement effective mechanism to address any violation of the consumers' sensitive data by placing adequate control to ensure security of personal data.

Despite the fact that all Web vendors today employ both the fair information practices and security information practices in their online transactions, consumers do not fully understand as to how the actions undertaken by Web vendors ease their risk. This may be due to a significant difference in the public perceptions and expert assessment of technology related risks. In order to enhance Web vendors' reputation, organizations should offer education and awareness programs on the efficiency of the protection mechanisms for sharing consumers' personal data online.

## VI. Limitations of the Study

The study has several limitations that affect the reliability and validity of the findings. The study did not take into account gender biases, cultural biases, income and other demographic variables with the research hypotheses. Further, only selected respondents participated in the study and therefore a self-selection bias might have affected the findings of this study and it may also limit the generalizability of the findings. Since sampling was based on convenience sample of students, there are chances that the responses provided might not be the true reflection of the population in general and the findings may not represent Malaysian consumers as a whole; therefore, any generalization of the findings may not be 100% reliable. The model may have excluded other possible factors influencing the consumers' trust in e-commerce transactions (i.e., the study did not consider other beliefs, such as perceived usefulness and perceived ease of use).

Future studies can also link other demographic variables of consumers as well as Web vendors' reputation, site's usefulness and ease of use. These dimensions may provide interesting recommendations on the difference in the consumers' trust building mechanisms to be adopted. Further, future studies can also differentiate between the perceptions of consumers who have not transacted online with the perceptions of consumers who have transacted online.

## VII. Conclusions

This study concludes that while trustworthiness of Web vendors is a critical factor in explaining consumers' trust to adopt e-commerce, it is important to pay attention to the consumers' risk concern on e-commerce transactions. Though in previous researches, security and privacy appear to be the top main concerns for consumers' trust in e-commerce adoption, the empirical results indicate that there is a poor correlation between perceived security and perceived privacy with consumers' trust. This may be because consumers get used to the Internet and to the techniques that can be used to protect themselves online, the security and privacy are becoming less sensitive matters over as time. However, the construct of perceived privacy manifests itself primarily through perceived security. As trustworthiness of Web Vendors lies at the heart of enduring B2C e-commerce relationship, web-based organizations need to find ways of improving consumers' perception of their trustworthiness in order to utilize fully the prospective of e-commerce.

## References

[1] Krishnan, G., "Internet marketing exposure in Malaysia," http://www.gobalakrishnan.com/2006/12/malaysia-internet-marketing/, 2006.

[2] Ahuja, M., Gupta, B. and Raman, P., "An Empirical investigation of online consumer purchasing behavior," Communications of the ACM, vol. 46, no. 12, pp. 145-151, 2003.

[3] Basu, A. and Muylle, S., "Authentication in e-commerce," Communications of the ACM, vol. 46, no. 12, pp. 159-166, 2003.

[4] Bingi, P., Mir, A. and Khamalah, J., "The challenges facing global e-commerce," Information System Management, vol. 17, no. 4, pp.26-34, 2000.

[5] Salam, A.F., Rao, H.R. and Pegels, C.C., "Consumer-perceived risk in e-commerce transactions", *Communications of the ACM*, vol. 46, no. 12, pp. 325-331, 2003.

[6] Ahmed, M., Hussein, R., Minakhatun, R. and Islam, R., "Building consumers' confidence in adopting e-commerce: a Malaysian case," Int. J. Business and Systems Research, vol. 1, no. 2, pp.236–255, 2007.

[7] Dauda, Y., Santhapparaj, AS., Asirvatham, D. and Raman, M., "The Impact of E-Commerce Security, and National Environment on Consumer adoption of Internet Banking in Malaysia and Singapore," *Journal of Internet Banking and Commerce*, vol. 12, no. 2, Aug 2007.

[8] Khatibi, A. Thyagarajan, V. and Scetharaman, A., "E-commerce in Malaysia: Perceived benefits and barriers", *Vikalpa*, Vol. 28, no.3, pp. 77-82, 2003.

[9] Economist Intelligence Unit, "Overview of e-commerce in Malaysia," *The Economist*, http://globaltechforum.eiu.com/index.asp?layout=printer_friendly&doc_id =8706, 13 June 2006.

[10] IDC Malaysia, "IDC Reports 70% Growth in Malaysia eCommerce Spending in 2006," http://www.idc.com.my/PressFiles/IDC%20Malaysia%20-%20eCommerce.asp, 24 January, 2007

[11] Jawahitha, S., "Consumer Protection in E-Commerce: Analysing the Statutes in Malaysia," *The Journal of American Academy of Business, Cambridge*.Vol. 4, no.1/2, pp. 55-63, 2004.

[12] Azmi, I.M., "E-commerce and privacy issues: an analysis of the personal data protection bill," International Review of Law Computers and Technology, vol. 16, no. 3, pp.317–330, 2002.

## Authors Profile

**Yi Yi Thaw** (rashidah_minakhatun@utp.edu.my) is a PhD student at the Department of Computer and Information Sciences, Universiti Teknologi PETRONAS, 31750, Tronoh, Perak, Malaysia.

**Dr. Ahmad Kamil Mahmood** (kamilmh@petronas.com.my) is an Associate Professor at the Department of Computer and Information Sciences, Universiti Teknologi PETRONAS, 31750, Tronoh, Perak, Malaysia.

**Dr. P.Dhanapal Durai Dominic** (dhanapal_d@petronas.com.my) is a Senior Lecturer at the Department of Computer and Information Sciences, Universiti Teknologi PETRONAS, 31750, Tronoh, Perak, Malaysia.